\documentclass[12pt]{article2}

\usepackage{scicite,subfigure,epsfig,amssymb,amsmath}

\usepackage{times}

\newenvironment{sciabstract}{%
\begin{quote} \bf}
{\end{quote}}

\renewcommand\refname{References and Notes}

\newcounter{lastnote}
\newenvironment{scilastnote}{%
\setcounter{lastnote}{\value{enumiv}}%
\addtocounter{lastnote}{+1}%
\begin{list}%
{\arabic{lastnote}.}
{\setlength{\leftmargin}{.22in}}
{\setlength{\labelsep}{.5em}}}
{\end{list}}

\title{Reassessing the Source of Long-Period Comets}

\author
{Nathan A. Kaib$^{1\ast}$ \& Thomas Quinn$^{1}$\\
\\
\normalsize{$^{1}$Department of Astronomy, University of Washington,}\\
\normalsize{Box 351580, U.W., Seattle, WA 98195-1580, USA}\\
\\
\normalsize{$^\ast$To whom correspondence should be addressed; E-mail:  kaib@astro.washington.edu.}
}

\date{}

\begin{document} 


\baselineskip24pt


\maketitle


\begin{sciabstract}
We present numerical simulations to model the production of observable long--period comets (LPCs) from the Oort Cloud, a vast reservoir of icy bodies surrounding the Sun.  We show that inner Oort Cloud objects can penetrate Jupiter's orbit via a largely unexplored dynamical pathway, and they are a significant, if not the dominant, source of known LPCs.  We use this LPC production to place observationally motivated constraints on the population and mass of the inner Oort Cloud, which are consistent with giant planet formation theory.  These constraints indicate that only one comet shower producing late Eocene bombardment levels has likely occurred since the Cambrian Explosion, making these phenomena an improbable cause of additional extinction events.
\end{sciabstract}

Because of their large distances from the Sun, Oort Cloud bodies undergo orbital evolution driven by gravitational perturbations from passing stars and the galactic tide\cite{oort50,heistre86}.  During this process, semimajor axes ($a$) remain nearly constant while perihelia (closest approach distances to the Sun, or $q$) evolve under these perturbations.  While most Oort Cloud bodies have perihelia far outside the planetary region of the Solar System, a tiny fraction are continually injected into planet-crossing orbits\cite{wietre99}.  Once there, they receive energy kicks from planetary perturbations during each perihelion passage, causing the semimajor axes to change at random.  The magnitude of these kicks increases greatly near Jupiter and Saturn, and in the inner 10--15 AU of the Solar System the typical planetary energy kick is greater than the gravitational binding energy of Oort Cloud orbits\cite{fern81}.  Thus, inward perihelion drift of LPCs is halted inside $\sim$10-15 AU, as they are either immediately ejected to interstellar space or perturbed onto low--$a$ (and therefore fixed--$q$) orbits before eventually being ejected.  This effect, known as the Jupiter-Saturn barrier, prevents many Oort Cloud bodies from passing near Earth.

To reach an observable orbit ($q <$ 5 AU), an Oort Cloud body's perihelion must therefore decrease from outside 10-15 AU to inside 5 AU in less than one orbital period.  Bodies with $a\lesssim$ 20,000 AU are less sensitive to galactic perturbations, and except during rare close stellar passages causing comet showers, their perihelia evolve too slowly to reach observable orbits before ejection.  For this reason, the Jupiter-Saturn barrier has traditionally been thought to prevent bodies in the inner 20,000 AU of the Oort Cloud from evolving to currently observable LPCs\cite{heistre86}, and this has been the motivation for dividing the Oort Cloud into the inner (unobservable) cloud with $a <$ 20,000 AU and outer (observable) cloud with $a >$ 20,000 AU.  The edge of the Jupiter-Saturn barrier is not abrupt, however, and recent modeling of scattered disk orbits indicates that repeated smaller planetary energy kicks near the barrier edge can inflate low--$a$ orbits to $a >$ 20,000 AU\cite{lev06}.  

Here, we present numerical simulations of the production of observable LPCs\cite{SOM}, and we find that many observable LPCs ultimately originate from the inner Oort Cloud (fig. 1).  In the case shown here, an inner Oort Cloud body with an initial semimajor axis of 6,000 AU had a perihelion slowly evolving Sunward under the influence of the galactic tide.  While the perihelion was beyond $\sim$18 AU, the semimajor axis was nearly unaltered by the relatively weak perturbations of Uranus and Neptune.  Between 18 and 14 AU, however, $a$ was rapidly inflated to $\sim$30,000 AU.  With such a large semimajor axis, the perihelion then decreased by 13 AU during the next orbital period, circumventing the Jupiter-Saturn barrier and evolving to an observable LPC.  About 85\% of inner Oort Cloud bodies evolving to observable LPCs followed an evolution similar to that shown in fig. 1.  Details of other known minor LPC production pathways that are less direct and efficient\cite{bras08a} can be found in the online material.

Although LPC production has been modeled previously, a substantial inner Oort Cloud contribution has hitherto gone unnoticed.  In some models, this is because planetary perturbations are not included\cite{heis87,rick08}, leaving no mechanism to inflate $a$ near the Jupiter-Saturn barrier.  However, some works include the full gravity of the giant planets\cite{wietre99,emel07}.  In these instances, orbital elements of incoming LPCs are sampled when they have already attained $q <$ 5 AU, obscuring any prior $a$--evolution.  We performed a similar orbit sampling of dynamically new LPCs (LPCs entering the observable region for the first time) (fig. 2a). The resulting $a$--distribution hides the inner Oort Cloud's contribution.

To assess the inner Oort Cloud's LPC production, we instead considered the initial semimajor axes from which LPCs evolved in our simulation (fig. 2a).  This distribution indicates that just over half of all dynamically new LPCs come from the inner Oort Cloud, which is a conservative estimate because we used an Oort Cloud model with a low inner-to-outer population ratio of 1.5:1 (see online material for a treatment of alternative models).  Below $a\sim$5,000 AU, LPC production fell dramatically because of the non-isotropic nature of the innermost part of our Oort Cloud model (fewer orbits had decreasing perihelia in this region).  However, even additional simulations with a more isotropized inner Oort Cloud\cite{kaibquinn08} indicate greatly suppressed LPC production below $a\sim$ 3,000 AU.  Thus, many bodies may reside inside $\sim$3,000 AU without producing observable LPCs (see online material), but any population with $a\gtrsim$ 3,000 AU will be reflected and constrained by the observed LPC population.

Because detecting LPCs outside the Jupiter-Saturn barrier is difficult, very few candidate inner Oort Cloud objects are currently known\cite{brown04,kaib09}.  However, as shown in Fig. 2a, our simulations predict that a substantial fraction of the hundreds of cataloged LPCs come from the inner Oort Cloud.  By assuming the inner Oort Cloud generates the entire observed LPC flux, we can estimate an upper limit on the number of cometary bodies with $a \gtrsim$ 3,000 AU.  Although the measured flux of LPCs is uncertain, even the highest flux values quoted in the literature\cite{ever67} imply an inner Oort Cloud population below $\sim$10$^{12}$, a factor of 2-3 greater than previously accepted estimates of the outer Oort Cloud population\cite{dones04}.  

Efficient LPC production from inner Oort Cloud orbits may also help resolve a possible inconsistency between the Oort Cloud mass and planet formation.  Because the outer Oort Cloud traps only 1-2\% of planetesimals scattered by giant planets\cite{kaibquinn08}, outer Oort Cloud population estimates imply an unreasonably massive primordial solar nebula for many assumed LPC population models\cite{dun08}.  While LPC population uncertainties may explain this discrepancy (see online material), inner Oort Cloud LPC production offers an alternative explanation.  This is because the inner Oort Cloud's trapping efficiency substantially exceeds that of the outer cloud if the Sun formed in a star cluster\cite{bras06,kaibquinn08}.  We predicted a new range of solar nebula masses as a function of inner Oort Cloud trapping efficiency (fig. 2b), assuming all known LPCs come from this region.  Higher trapping efficiencies (5-10\%) provided by a solar birth cluster imply a solar nebula mass range (20-100 M$_{\earth}$ in solids) compatible with giant planet formation theory\cite{gom05,thom03}.  This raises the possibility that known LPCs can be used to discriminate between solar birth environments\cite{SOM}. 

With an upper estimate on the population between 3,000 AU $<a<$ 20,000 AU, we can now also determine the maximum number of LPCs delivered during comet showers because this region supplies the excess LPCs absent during non--shower periods\cite{hills81}.  From numerical experiments\cite{SOM} of individual stellar encounters (an example of which is shown in fig. 3a), we find that the number of comets injected into Earth-crossing orbits scales with the Sun's velocity change due to the encounter (fig. 3b).  Using a stellar population model for the solar neighborhood\cite{rick08}, we transformed this into a prediction of comet shower strength vs. shower frequency (fig. 3c).  By attributing the observed Earth-crossing LPC flux estimate [two dynamically new LPCs/yr\cite{heis87}] to just the inner Oort Cloud,  we predicted a maximum impact number\cite{weiss07} expected for a given comet shower (fig. 3c).  

Spikes in the flux of Earth-impacting LPCs during comet showers have previously been suggested as a possible cause of mass extinctions seen in the fossil record\cite{hut87}.  With 3 nearly simultaneous major impacts accompanied by a 2.5-Myr $^{3}$He spike, a comet shower is a particularly appealing explanation for the late Eocene extinction\cite{far98}.   On the basis of our upper estimate of the inner Oort Cloud population, it is expected that the most powerful comet shower in the past 500 Myrs yielded just 2-3 impacting comets greater than $\sim$2 km in diameter (fig. 3c).  If the late Eocene episode was caused by a comet shower, it was likely the most powerful shower since the Cambrian Explosion, implying that comet showers are unlikely to account for other observed extinction events.  Semimajor axis distributions much more extreme than our chosen Oort Cloud model can raise the shower intensity by a factor of $\sim3$, but solar nebula mass requirements to form these configurations make them physically implausible (see online material for details).

While known LPCs constrain the total population beyond $a\sim$ 3,000 AU, they offer little information about the relative inner and outer Oort Cloud populations because LPC production obscures orbital histories.  However, inner cloud LPC production predicts the generation of $a>$20,000 AU orbits near Jupiter and Saturn.  In contrast, the outer Oort Cloud's $a>$20,000 AU population will decreaese after passing through this region..  Thus, a comparison of original and future semimajor axes for a large LPC sample near $q\sim$ 10 AU (analogous to that done for currently known LPCs) could provide an opportunity to distinguish between production mechanisms.  

\begin{figure}[p]
\centering
\includegraphics[scale=1.5]{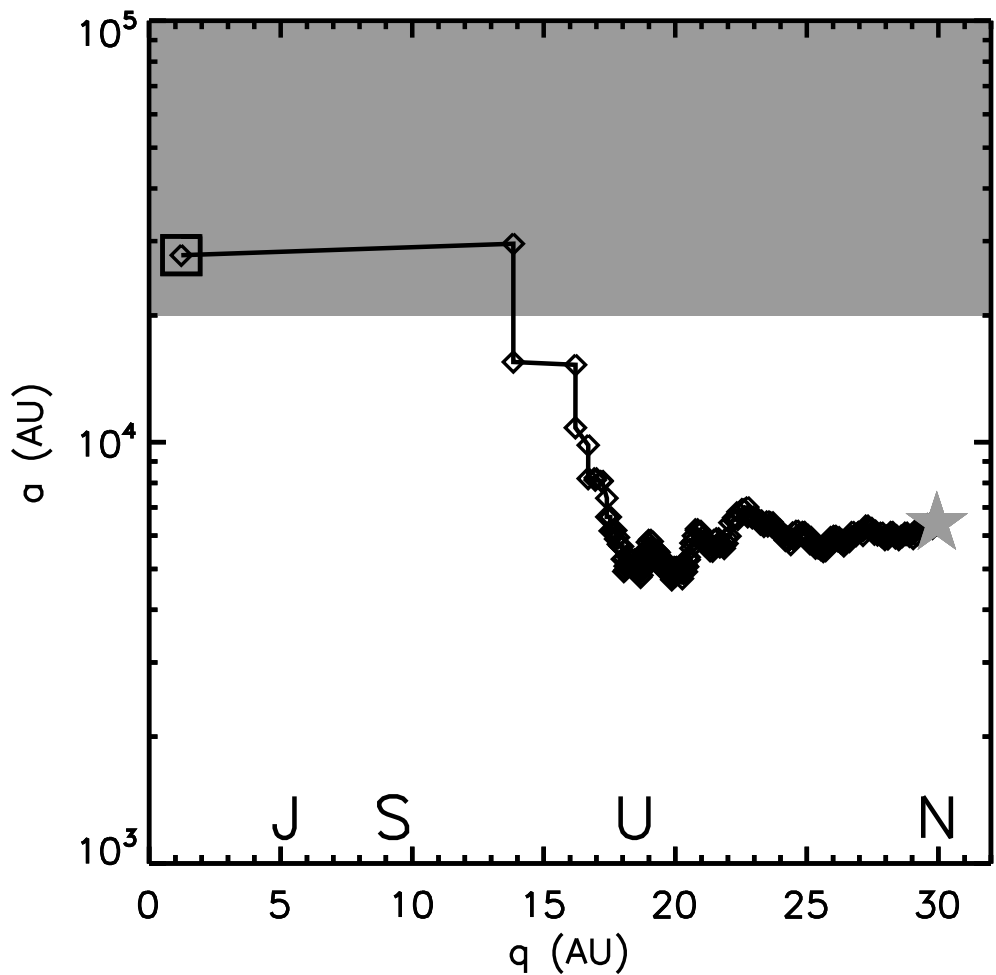}
\caption{An example of the typical evolution from an inner Oort Cloud object into an observable LPC.  Orbital elements are sampled each time the object crosses the $r = 35$ AU boundary (twice per orbit).  The star data point marks the start of the evolution and the square marks the end.  The shaded area indicates the $a$-range of the outer Oort Cloud, and the perihelia of the giant planets are noted using their initials.}
\end{figure}

\begin{figure}[p]
\centering
\includegraphics[scale=.9]{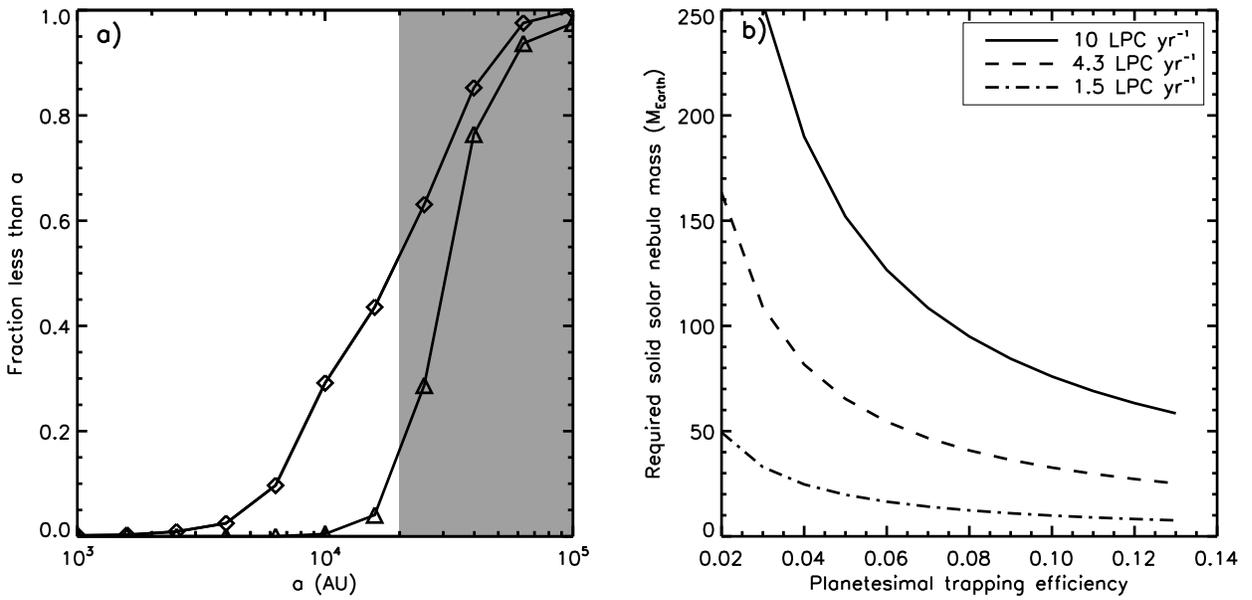}
\caption{{\it\bf a:} Cumulative semimajor axis distributions for dynamically new LPCs.  The two distributions are the initial ($t=0$) values ({\it diamonds}) and when $q$ first drops below 5 AU ({\it triangles}). {\it\bf b:} The required solar nebula mass in solids (adopting an average LPC mass of 4 x 10$^{16}$ g as a fiducial value\cite{weiss96}) as a function of inner Oort Cloud trapping efficiency for an assumed dynamically new LPC flux.  The different curves correspond to flux values of 10\cite{ever67}, 4.3\cite{fran05}, and 1.5\cite{nes07} dynamically new LPCs/yr.}
\end{figure}

\begin{figure}[p]
\centering
\includegraphics[scale=.9]{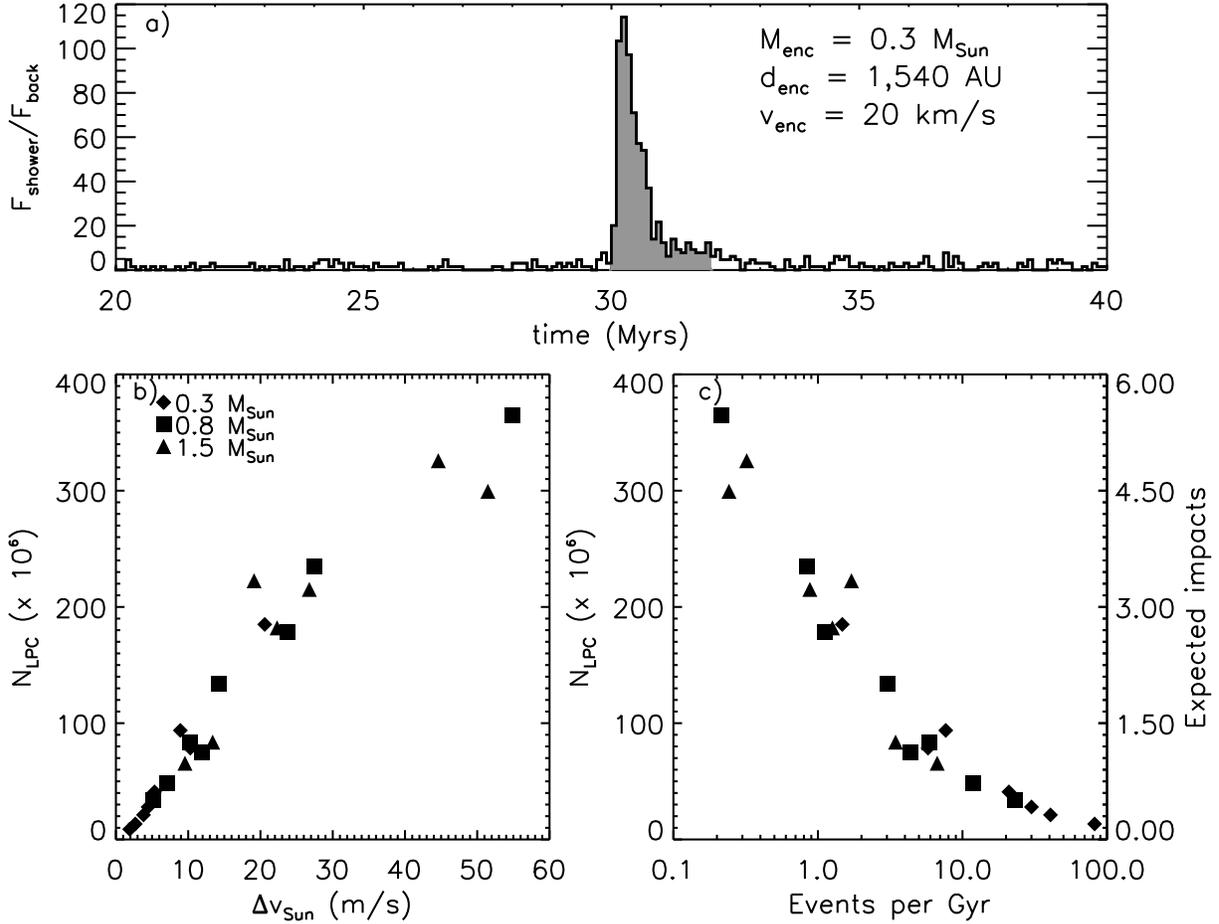}
\caption{{\it\bf a:} Flux of dynamically new LPCs (in units of the average background level) vs. time during a comet shower at $t = 30$ Myrs.  To quantify ``comet shower strength'' all dynamically new LPCs in the 2 Myrs following the stellar passage are counted (shaded region).  {\it\bf b:}  The number of Earth-crossing LPCs delivered during comet showers vs. the solar velocity change from the stellar encounter (assuming equal scaling of observable and Earth-crossing LPCs).  The different symbols refer to different stellar encounter masses ($M_{enc}$).  Encounter distances ($d_{enc}$) ranged from 1,300 AU to 7,000 AU, and velocities ($v_{enc}$) were between 20 and 40 km/s.  {\it\bf c:}  The number of Earth-crossing comets delivered in comet showers vs. the expected shower frequency at that strength.  On the right y-axis the expected number of terrestrial impacts with a comet greater than $\sim$2 km in diameter is shown.}
\end{figure}

\section{Supporting Online Material}

\subsection{Materials and Methods}

\subsubsection{Background LPC Production Simulations}

In our simulations modeling LPC production, we begin with the Sun and the four giant planets on their current orbits.  In addition, we include $\sim$10$^6$ massless test bodies representing Oort Cloud bodies.  Initial semimajor axis distributions are drawn from the end of previous 4.5-Gyr Oort Cloud formation simulations\cite{kaibquinn08}, while different $e$, $i$, $\Omega$, and $\omega$ distributions are applied for different ranges of $a$ to simulate varying levels of cloud isotropy.  Two sets of initial conditions are used.  The first, consisting of 10$^6$ particles, has an istropic outer cloud and a less isotropized inner cloud.  We refer to this as the ``classic'' model because it corresponds to Oort Cloud formation within the Sun's current galactic setting.  The second set of initial conditions, called the ``cluster'' model, has a more isotropized and enriched inner cloud produced by the Sun's initial residence in a putative birth cluster, which in this case is an open cluster with a mean density of 30 stars/pc$^3$\cite{kaibquinn08}.  For the cluster simulation we only use 500,000 test particles.  

The two simulations' initial conditions are compared in Figure S1.  In Figure S1a, we see that beyond $a > 10^4$ AU the mean cosine of inclination is very close to zero for both distributions, indicative of random inclination distributions.  However, the mean cosine for the classic model rises to $\sim$0.6 between 10$^4$ AU and 6000 AU indicating a non-isotropic innermost edge of the Oort Cloud that is concentrated toward the ecliptic plane.  In this region, stellar and galactic tidal perturbations are too weak to completely randomize Oort Cloud orbits.  There is a similar but less dramatic rise to $\sim$0.3 for the cluster distribution in the same $a$-range.  This is because an istropic inner cloud was formed by the powerful perturbations of an open cluster environment, but Oort Cloud formation continued well after cluster dispersal, superimposing a non-isotropic classic cloud on the isotropized population\cite{kaibquinn08}.

In Figure S1b, the semimajor axis distributions for the two comet clouds are also compared.  With a 3.5:1 inner-to-outer cloud population ratio, the cluster distribution is more centrally concentrated than the classic, which only has a 1.5:1 ratio (defining the inner cloud as $a <$ 20,000 AU).  Once again, the increased central concentration of the cluster distribution is due to the effects of the early star cluster whose powerful perturbations pull objects into the Oort Cloud more efficiently at low $a$\cite{fern97,bras06,kaibquinn08}.

Once we generate our initial particle orbits, we evolve the test particles under the influence of the galactic tide and passing stars for 1.2 Gyrs.  Our galactic tidal model\cite{lev01} assumes a dominant vertical component due to a solar neighborhood disk density of 0.1 M$_{\sun}$/pc$^3$ as well as a radial tidal component that is $\sim$1 order of magnitude smaller than the vertical.  In addition to perturbations from the galactic tide, our stellar passage model\cite{rick08} generates 10.8 stellar passages within a parsec of the Sun every Myr on average.

To integrate our particles we use an N-body code based on SWIFT\cite{levdun94} that is enhanced with variable timestepping to maximize computing efficiency\cite{kaib09}.  This heirarchical timestepping algorithm takes 100-day timesteps inside 300 AU, 10-year steps between 300 and 400 AU, and 25-year steps outside 400 AU \cite{kaib09}.  It should be noted that massive bodies are always integrated with a 100-day timestep.  When integrating test particles using large timesteps, the planetary integrations are performed first, allowing us to use their initial and final positions for the test particle integration.  While variable timestepping greatly increases our code's speed, it also increases integration errors resulting in an artificial random walk of orbital elements.  However, this error-driven random walk is less than 1\% of the magnitude of the random walks due to real gravitational perturbations and does not impact our final results\cite{kaibquinn08}.  

Lastly, to minimize any effects from our choice of initial conditions, we only analyze the last 200 Myrs of simulations.  During this final analysis, we turn off stellar encounters.  This is done to isolate the LPC production mechanism described in the main text and avoid the ``pollution'' of LPCs that are directly injected from the inner Oort Cloud during particularly strong stellar passages\cite{emel07}.  Additionally, there is no evidence that the Sun has recently undergone a close stellar encounter\cite{weiss96,garc01}, so comparing a simulated LPC flux influenced by such events to the real LPC flux may be inaccurate.  The exclusion of even weak stellar encounters from our analysis window may be an excessive measure, however.  To assess the significance of this choice, we also integrate the last 200 Myrs of our classic simulation with stellar encounters.  During this integration, there is a 60-Myr window without a stellar passage inside 10$^4$ AU that we can use to analyze.  During this window, the inner and outer Oort Cloud LPC production rates only increase by 12\% and 4\% respectively compared to our simulation without encounters.  Thus, the conclusions of the main text are essentially unaltered by the effects of weak stellar encounters.  

\subsubsection{Comet Shower Simulations}

To generate initial conditions for our short simulations of comet showers, we utilize the large steady-state comet clouds that we have built in our 1.2 Gyr runs.  To build a comet cloud for a short run, we randomly select particles after $t=$ 1 Gyr from our long run cloud until a 10$^6$ particle Oort Cloud is assembled.  For each particle we shift its original orbital elements by a random value between -0.01\% and 0.01\%, so that all of our initial orbits are slightly different in each simulation.  These clouds are then evolved for 30 Myrs to minimize any effects of our initial orbital element shifting before a comet shower is triggered at $t = 30$ Myrs by a stellar encounter with a specified mass, velocity, and encounter.  Stellar passages are started at an initial distance of 200,000 AU with the particular angular orientation of the passage randomly selected.  Each simulation is run until $t = 40$ Myrs.

\subsection{Results and Discussion}
\subsubsection{LPC Production}
As stated in the main text, we find that the inner Oort Cloud is a major source of observable LPC production in both of our simulations.  In our classic simulation, we find that 51\% of dynamically new LPCs entering the terrestrial region ($q <$ 5 AU) were initially placed on orbits with $a <$ 20,000 AU.  In the cluster simulation this percentage of LPCs from the inner Oort Cloud jumps up to 67\%.  Although the cluster simulation has an inner-to-outer cloud ratio over twice as large as the classic, the increase in the inner cloud's LPC production is more modest.  The reason for this is that many of the additional inner Oort Cloud comets are found on orbits with $a <$ 3,000 AU.  In this orbital regime, comets are more insulated from the effects of galactic tidal and stellar perturbations, leading to a lower flux of bodies into the planetary region.  Furthermore, the greater range in orbital energy traversed by potential LPCs at this $a$-range before encountering the Jupiter-Saturn barrier appears to suppress comet production.  Hence, there seems to be an optimum range for inner Oort Cloud LPC production from 3,000 to 20,000 AU.  When we only consider semimajor axes beyond 3,000 AU, the inner Oort Cloud comprises 56\% of the classic cloud's comets and 72\% of the cluster cloud's comets.  Considering these revised figures, the inner cloud's LPC contribution in each simulation scales linearly with population.

As comets are injected back into the planetary region ($q <$ 35 AU), outer Oort Cloud bodies are more likely than inner cloud bodies to reach the terrestrial planetary region ($q <$ 5 AU) before being ejected.  During the course of our integrations, one in every 11,500 outer Oort Cloud bodies enters the planetary region as a dynamically new body every Myr.  For the terrestrial planet region, this fraction drops by a factor of $\sim$6 to one in every 71,000 outer Oort Cloud bodies per Myr.  While the inner Oort Cloud injects a similar fraction of bodies into the planetary region (1 in 9,700 bodies per Myr), only 1/12 of these bodies reach the terrestrial planet region resulting in a dynamically new LPC for every 116,000 inner Oort Cloud bodies per Myr.  Thus, a planetary interloper from the inner Oort Cloud is only half as likely to evolve to an observable LPC as one from the outer Oort Cloud.

In addition to the production mechanism outlined in Figure 1 of the main text, $\sim$15\% of LPCs from the inner Oort Cloud reach the terrestrial planet region via two other known pathways\cite{bras08a}.  The first and most common alternative pathway is shown in Figure S2a.  This evolution accounts for $\sim$12\% of LPCs from the inner Oort Cloud.  Here an inner Oort Cloud object's initial semimajor axis of $\sim$17,000 AU remains essentially unchanged until a major stellar encounter at $t=$ 730 Myrs.  In addition to altering perihelia, stellar encounters can also produce semimajor axis changes (although less efficiently), and this encounter inflates the semimajor axis to over 30,000 AU.  Later, at $t=$ 1.1 Gyrs, the object enters the terrestrial planet region as an LPC with $a >$ 20,000 AU.  This pathway accounts for the majority of inner Oort Cloud LPCs that do not evolve according to Figure 1 of the main text.  Because semimajor axis diffusion due to stellar encounters is relatively inefficient\cite{bras08a, kaibquinn08}, most of the LPCs generated by this mechanism have initial semimajor axes between 15,000 AU and 20,000 AU.

A third possible evolutionary path from the inner Oort Cloud to the terrestrial planet region also exists.  As stated previously, 11 in 12 comets from the inner Oort Cloud that enter the planetary region fail to become observable LPCs.   Of these,  26.2\% are ejected from the solar system, while 51.4\% return to the inner Oort Cloud.  The remaining 22.4\% are transferred to outer Oort Cloud orbits via planetary energy kicks.  An example of such an object is shown in Figure S2b.  In this particular example, a comet with an initial semimajor axis of 10,000 AU has its perihelion injected into planetary region at $t =$ 290 Myrs.  After receiving planetary energy kicks for $\sim$100 Myrs, this object's perihelion evolves back out of the planetary region with the semimajor axis inflated to over 40,000 AU.  After further $a$-evolution due to stellar encounters, the object is injected into the terrestrial planet region as an LPC with $a >$ 30,000 AU.  Although this pathway accounts for just $\sim$3\% of inner Oort Cloud LPCs, this mechanism tends to generate LPCs from lower initial semimajor axes compared to the pathway in Figure S2a since planetary perturbations can alter semimajor axes more efficiently than stellar encounters.

\subsubsection{LPC Orbits}

In Figure S3, we show the distribution of orbital inclinations for observable LPCs produced in our simulations.  While statistically consistent with an isotropic distribution, there is a slight bias toward retrograde orbits (55\%).  This bias is present for comets originating from both the inner and outer Oort Clouds.  The reason for this is because retrograde orbits experience weaker planetary perturbations on average compared to prograde orbits.  As a result, the loss cylinder due to the Jupiter-Saturn barrier is smaller for retrograde orbits, enabling lower semimajor axes to penetrate the terrestrial planet region than for prograde orbits.  There is also an observational bias toward the discovery of retrograde LPC orbits\cite{ever67}, but because the selection effects of LPC observations are so difficult to characterize, it is possible that the actual bias predicted by our simulations may be small enough to have gone unnoticed.

It is somewhat surprising that our LPC sample shows a slight bias against prograde orbits when the inner Oort Cloud has a substantial bias toward low inclinations.  Although the entire population inside $a <$ 10,000 AU displays a prograde bias, the bodies that actually evolve to observable LPCs are an isotropized subpopulation.  These are bodies that have either been strongly perturbed by a stellar encounter or have undergone a full Kozai tidal cycle.  Because the galactic tide causes a rapid precession of the longitude of ascending node in the galactic frame (which produces a precession of the ecliptic inclination), ecliptic inclinations are istropized before perihelia complete a full oscillation cycle for this range of $a$\cite{lev06}.

In Figure S4, we also examine the semimajor axis distribution of observable LPCs as they enter the terrestrial planet region.  When comparing LPCs originating from the inner Oort Cloud vs. the outer Oort Cloud, the two distributions differ substantially.  LPCs originating from the outer cloud have a median semimajor axis of 35,000 AU, whereas those from the inner Oort Cloud have a lower median value of 26,000 AU.  This is to be expected since inner Oort Cloud bodies begin on semimajor axes too small to circumvent the Jupiter-Saturn barrier, and most will have their semimajor axes inflated to values just high enough to reach the terrestrial planet region.  On the other hand, outer Oort Cloud bodies undergo little semimajor axis modification and can take any value greater than $\sim$20,0000 AU.  This difference in distributions is compelling, for observed dynamically new comets have a median semimajor axis of 27,000 AU\cite{dones04}, which is very close to the median value of our inner Oort Cloud LPC distribution.  Moreover, models of LPC production from the outer Oort Cloud tend to predict a median semimajor axis of new LPCs that is too large\cite{heis87,wietre99}.  This may indicate that the inner Oort Cloud is in fact the dominant source of LPCs.  

\subsubsection{Inner Oort Cloud Population}

In the main text we calculate the maximum population of the inner Oort Cloud in our particular model if the entire LPC flux is produced by this region.  However, we also show that for $a<$ 3,000 AU, LPC production is much less efficient than for larger $a$.  Consequently, it is possible to hide many comets at $a<$ 3,000 AU without increasing the observed LPC flux.  Although it is very difficult for the stellar and tidal perturbations of the Sun's current galactic environment to populate the Oort Cloud at this $a$-range, it is quite possible to enrich this region if the Sun was born into a star cluster, for stronger gravitational perturbations are inherent to such environments\cite{fern97,bras06,kaibquinn08}.  For this reason, we repeat the maximum population calculations of the main text using alternative $a$-distributions of the inner Oort Cloud.  These maximum population extimates are shown in Figure S5.  From this figure, we see that for virtually any inner Oort Cloud model with a median semimajor axis above $\sim$2,000 AU, the total population must be below $\sim$2 x 10$^{12}$, only a factor of two greater than the value determined in the main text.  Inside $\sim$2,000 AU, much greater numbers of comets can exist, however.

In these calculations, we assume that the entire inner Oort Cloud is distributed over one decade of $a$-space since this is the size of typical comet clouds assembled in embedded clusters\cite{bras06}.  In addition, we use the results from our ``cluster'' simulation to evaluate LPC production levels at these low $a$-ranges, as the ``classic'' simulation contains a less isotropized inner Oort Cloud, and a cluster environment tends to produce a very isotropic orbital distribution\cite{kaibquinn08}.  Lastly, it should be noted that the density profile for our cluster simulation is proportional to $r^{-3.4}$, which is quite similar to most embedded cluster simulations.  Therefore, using smaller $a$-ranges of this cloud should accurately predict the LPC flux from comet clouds formed in embedded clusters.

\subsubsection{Protoplanetary Disk Mass and the Solar Birth Environment}

While the inner Oort Cloud does not produce observable LPCs as efficiently as the outer cloud, the production efficiency for the two regions differ by less than a factor of 2.  For this reason, it is possible that the inner Oort Cloud is the source of most observed LPCs.  As stated in the main text, this is advantageous because some estimates of the typical LPC mass require an unreasonably massive protoplanetary disk to form an outer Oort Cloud capable of producing the currently observed LPC flux.  It should be stressed that attributing LPCs to the inner Oort Cloud does not simply transfer this paradox to another region because the formation of a massive inner cloud is much more physically plausible\cite{fern97}.  Models of Oort Cloud formation inside a putative solar birth cluster environment show that powerful cluster perturbations trap planetesimals in the inner Oort Cloud at a rate 2-10 times greater than the outer cloud\cite{bras06,kaibquinn08}.  

Thus, LPC production from an inner Oort Cloud whose mass has been enhanced by an early star cluster environment may explain any perceived Oort Cloud mass problem.  Also in Figure S5, we calculate the primordial solar nebula mass in solids required between 4 and 40 AU to generate the maximum inner Oort Cloud populations plotted.  For these calculations, we assume an Oort Cloud trapping efficiency of 10\%, as this is typical for an embedded cluster environment.  Figure S5 shows that to generate all observed LPCs from a comet cloud formed in an embedded cluster, clouds with median semimajor axes below 2,000 AU require exceptionally high solar nebula masses ($\gtrsim200$ M$_{\earth}$).  In reality, the Oort Cloud is probably some combination of a massive cluster-style cloud and a less massive classic distribution\cite{kaibquinn08,bras08}, so the mass requirements shown in Figure S5 should be considered upper estimates.

If the bulk of observed LPCs originate from an enriched inner Oort Cloud, these comets may provide a possible constraint on the Sun's primordial environment.  In Figure S5, we also show the general scaling found between mean embedded cluster density ($\rho$) and comet cloud semimajor axis\cite{bras06}.  Many denser cluster environments place the bulk of Oort Cloud bodies on orbital semimajor axes too small to be affected by the solar neighborhood's present set of stellar and tidal perturbers.  Although a great deal of mass is trapped in the inner Oort Cloud in these cases, they will remain locked in their primordial orbits and contribute no LPCs today.  Additionally, the orbit of Sedna\cite{brown04} provides another constraint on the Sun's early environment if this object's orbit was indeed sculpted by an early strong external pertrubation to the solar system\cite{morblev04,bras06,kaibquinn08}.  Therefore, the Sun's early environment must have provided powerful enough perturbations to produce Sedna at the Oort Cloud's inner boundary and also populate the ``sweet spot'' of inner Oort Cloud LPC production discussed in the previous section.  In the case of embedded star clusters\cite{bras06}, mean cluster densities $\gtrsim$10$^{3}$ M$_{\sun}$/pc$^3$ provide both a semimajor axis range (500 AU $<a<$ 10$^4$ AU) and a trapping efficiency ($\sim$10-12\%) capable of producing today's observed LPC flux along with the ability to reproduce Sedna.  Although open cluster environments are much more stochastic, previous modeling shows that they are also capable of providing similar trapping efficiencies and semimajor axis ranges\cite{kaibquinn08}.  

If the embedded cluster environment quoted above is used, then approximately 5-6\% of planetesimals scattered by the giant planets will be placed inside an $a$-range accessed by LPC production (3,000 AU $< a <$ 10$^4$ AU).  In the main text, we have shown that a population of 10$^{12}$ km-sized bodies in this region can explain an observed LPC flux near Earth of 2 dynamically new LPCs/yr.  Now, using an average comet mass as a fiducial value of 4 x 10$^{16}$ g\cite{weiss96} yields an Oort Cloud mass of 6.7 M$_{\earth}$ in this region and requires that 100-120 M$_{\earth}$ of solid material was in the form of comet-sized planetesimals in the original protoplanetary disk of the outer solar system.  However, it is quite likely that after cluster dispersal another $\sim$2\% of planetesimals are trapped in the inner Oort Cloud and another $\sim$1\% in the outer cloud\cite{kaibquinn08,bras08}.  This would lower the comet component of the protoplanetary disk mass to 65-75 M$_{\earth}$.  

\subsubsection{Mass Calculation Uncertainties}

It must be emphasized that all of the above mass calculations are dependent on the LPC flux and size distribution, both of which are poorly constrained\cite{dun08}.  Because only a handful of LPCs have had nuclei sizes reliably measured\cite{lamy04}, there is currently no way to directly measure a size distribution of these objects.  As a result, one must must estimate LPC nuclei sizes by employing a crude conversion of the measured total comet brightness\cite{weiss96}, which of course is dominated by the coma rather than the nucleus.  Following this size estimate, a mass is determined by assuming a comet density, typically .6 g/cm$^3$.

Once a mean comet mass is estimated, the total Oort Cloud mass is inferred from the observed flux of dynamically new LPCs, which is also fraught with uncertainty.  Until recently, most LPCs were found by amateur astronomers, making observational selection biases extremely difficult to characterize\cite{ever67}.  Even though many recent LPCs have been observed and discovered by the LINEAR survey rather than by amateur astronomer, these are difficult to characterize since their brightnesses are reported assuming they are point sources\cite{fran05}.  In addition, dynamically new LPCs are known to be consistently brighter than returning LPCs leading to differing discovery efficiencies for the two populations.  Estimates of how different the two discovery efficiencies are vary substantially\cite{nes07}.  Because not all LPCs entering the terrestrial planet region are discovered by astronomers, any LPC flux estimate will be quite sensitive to the estimated enhanced discovery efficiency of dynamically new LPCs.

For our calculations in Figure S5 and the main text, we have adopted a mean LPC mass of 4 x 10$^{16}$ g\cite{weiss96} as a fiducial value.  This figure is based on a specific assumed LPC magnitude distribution\cite{ever67a} and a conversion of comet brightness to nucleus size based on measurements of Halley's comet.  Other studies\cite{fran05,hugh01} have found shallower LPC magnitude distributions than the one we employ, which would imply a higher mean comet mass and Oort Cloud mass than our calculations.  However, due to a lack in observed small LPCs, these same studies also found dynamically new LPC flux values that were a factor of 2--6 lower\cite{fran05,nes07} than our assumed value of 10 yr${-1}$\cite{ever67}.  This would counteract the increased mean mass in our calculations, again leading to lower Oort Cloud mass estimates.  In addition, there exists an alternative LPC magnitude-to-mass relation\cite{bailstag88} from the one we employ, and this yields lower LPC masses than the one used in our calculations.  

Thus, the extreme requirements placed on the primordial solar nebula mass may simply be the result of the huge uncertainties in the LPC flux and size distribution.  Previous outer Oort Cloud mass estimates range between 2 and 60 M$_{\earth}$ (nearly 2 orders of magnitude!) with lower values being favored\cite{dun08}, since they are more consistent with planet formation theory.  Regardless of the LPC observational uncertainties, however, the comet production mechanism discussed in the main text shows that the Oort Cloud as a whole produces LPCs more efficiently than assumed in previous works, and higher LPC flux and mass values do not necessarily require an incredibly massive primordial solar nebula to account for the observed comet flux.

\subsubsection{Comet Showers}

In the main text, we estimate the number of impacts expected from the most powerful comet shower since the Cambrian Explosion.  To calculate this, we count the number of dynamically new LPCs generated from the inner Oort Cloud in the 2 Myrs following the shower-triggering stellar passage.  We then multiply this number by the ratio of the real Earth-crossing LPC flux (2.1 yr$^{-1}$) to the normal background LPC flux from the inner Oort Cloud seen in our simulation.  This then yields the number of dynamically new Earth-crossing LPCs expected from a given shower.  On average, a typical LPC has a 2.4 x 10$^{-9}$ probability of hitting the Earth per perihelion passage\cite{weiss07}.  Physical disruption models predict that typical LPCs make 4.4 perhelion passages on average before being disrupted or ejected\cite{weiss79}, and we are able to replicate this finding by applying this model to our background LPCs.  Because comet shower LPCs have lower initial semimajor axes than normal background LPCs, they are less susceptible to ejection by the planets.  While previous modeling of comet showers predicts these LPCs will make 8.6 perihelion passages on average\cite{hut87}, we find a mean of 6 perihelion passages when applying the same disruption model to the LPCs generated by our most powerful showers expected in the past 500 Myrs.  As a result, we multipy the above LPC impact probability by a factor of 6.  This number multiplied by the number of dynamically new LPCs from the inner Oort Cloud yields the predicted number of comet impacts.

The numbers calculated in the main text are derived from the classic Oort Cloud model, yet we have discussed previously how the Oort Cloud's distribution of comets can be greatly impacted by the Sun's dynamical history.  It is therefore worth exploring how much the results of the main text vary for alternative distributions of comets.  To do this we perform three additional simulations of the most powerful comet shower expected in the past 500 Myrs ($\Delta v_{\sun} =$ 17.5 m/s) and then combine the resulting shower LPCs to build a larger statistical comet sample.  Following this, we consider only half-decade ranges of $a$ and calculate the ratio of shower to non-shower LPC production for each $a$-range.  Using this result, we then determine the maximum number of impacts expected if the Oort Cloud was only found within this range.  We use the cluster simulation for all of this work since enriched specific regions of the Oort Cloud with a solar birth cluster also implies an isotropized inner Oort Cloud.  

Our results are shown in Figure S6.  As can be seen, the results presented in the main text are fairly robust.  The reason for this is that LPC production is reasonably efficient throughout 3,000 AU $<a<$ 20,000 AU, so the difference between shower LPC production and non-shower production can never be much higher in this $a$-range than what was found in the main text.  In constrast, LPC production is much less efficient inside 3,000 AU, and shower intensity increases by a factor of $\sim$3 compared to Figure 3 in the main text if the entire Oort Cloud is located within 3,000 AU.  However, Figure S6 assumes that this region's population is the maximum allowed by the observed LPC flux, and this would imply an Oort Cloud mass of 125 M$_{\earth}$ and (assuming a 10\% trapping efficiency) a primordial solar nebula mass in solids of 1250 M$_{\earth}$!  Thus, this model is probably not physically possible.  We can conclude then that the comet shower analysis in the main text is an acceptable estimate of shower intensity for most physically consistent Oort Cloud models.


\setcounter{figure}{0}
\renewcommand{\thefigure}{S\arabic{figure}}

\begin{figure}
\centering
\includegraphics{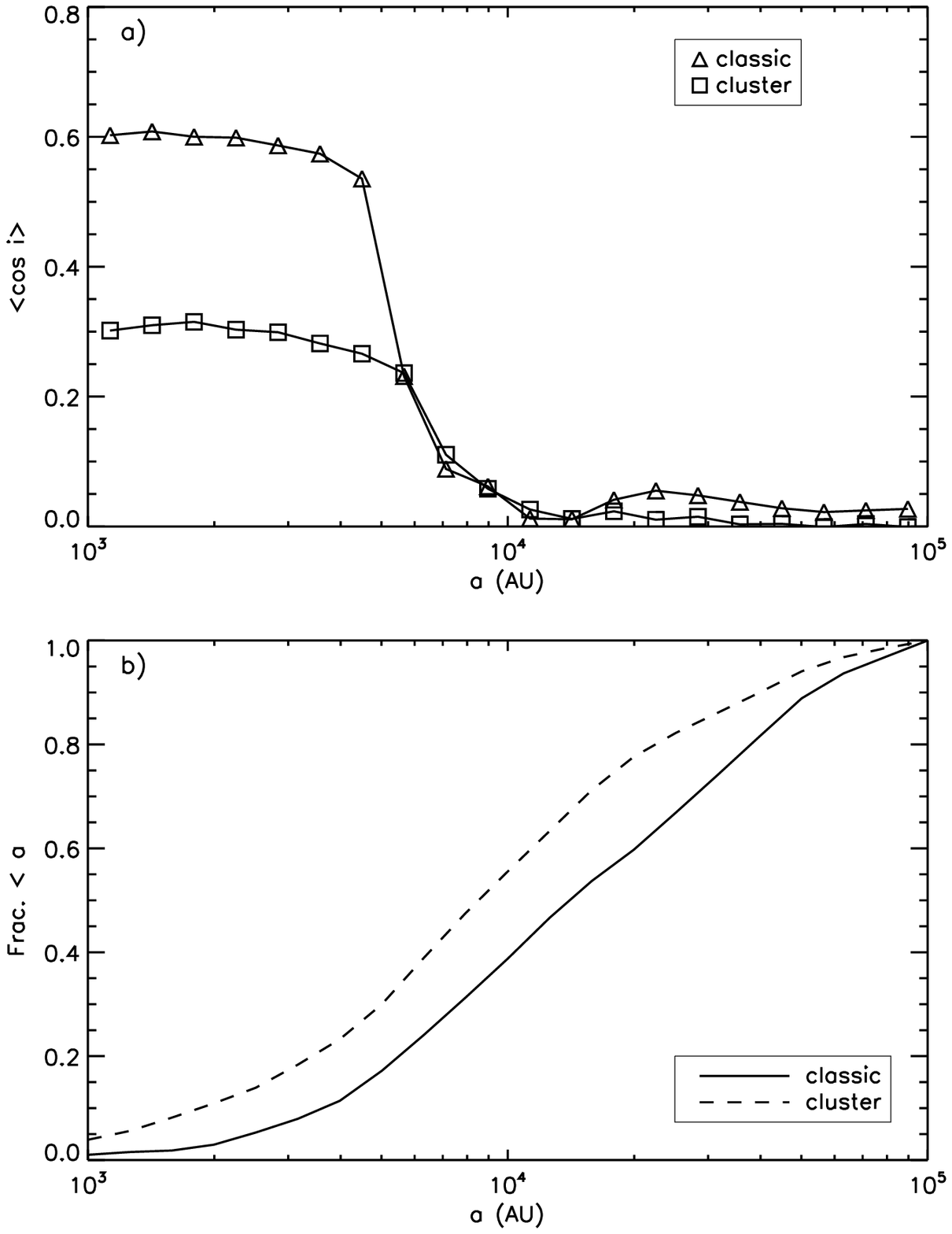}
\caption[S1]{{\it\bf a:} Comparison of the initial inclination distributions of the classic simulation (triangles) and the cluster simulation (squares).  The mean cosine of the inclination with respect to the ecliptic is plotted against semimajor axes for all bodies in the Oort Cloud. {\it\bf b:} Comparison of the initial cumulative semimajor axis distributions of the classic simulation (solid line) and the cluster simulation (dashed line).  }\label{fig:1}
\end{figure}

\begin{figure}
\centering
\includegraphics{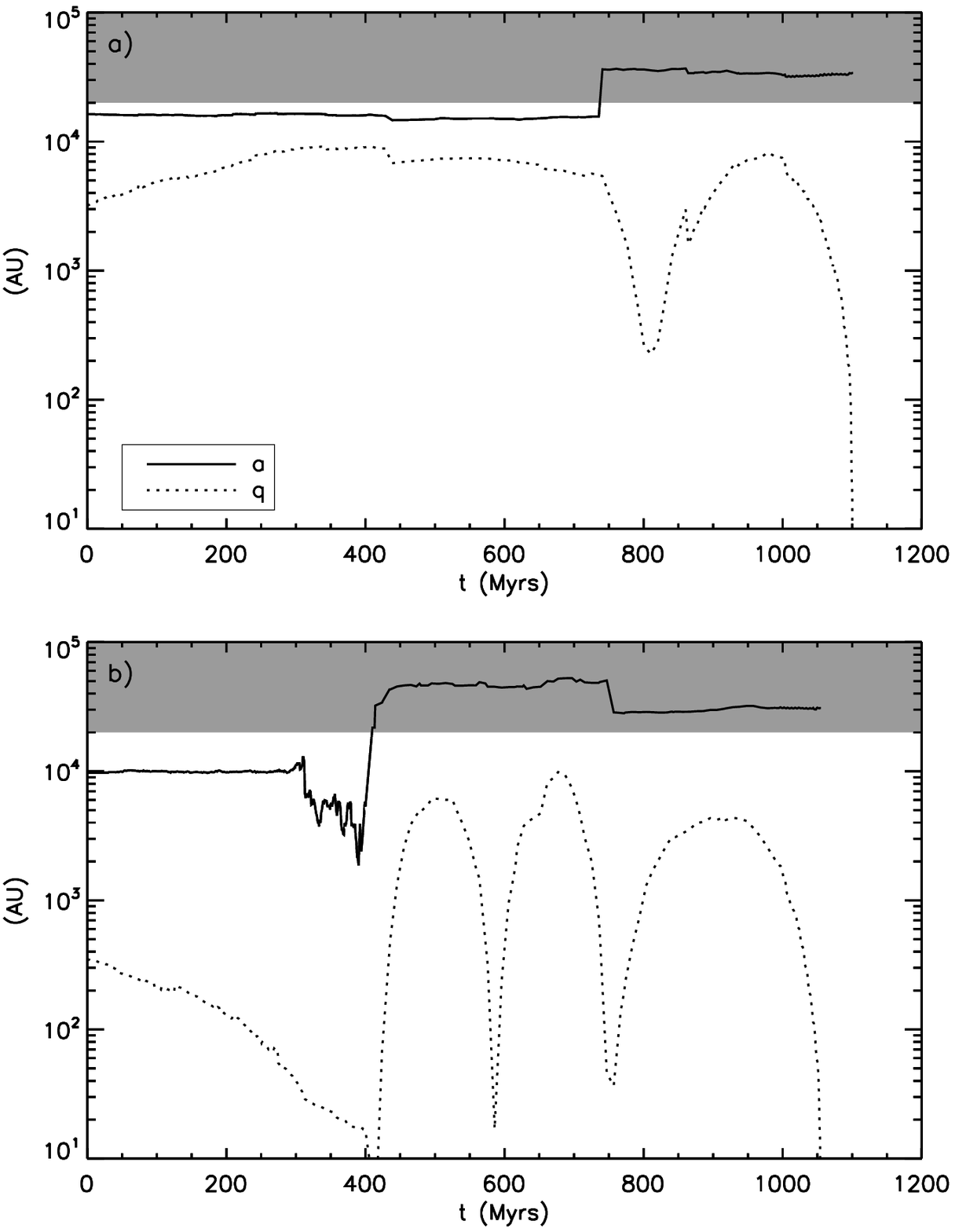}
\caption{Time evolution of two different orbits initially located in the inner Oort Cloud. Semimajor axis is shown with the solid line, while perihelion is marked by the dotted line.  The semimajor axis region of the outer Oort Cloud is marked by the shaded region.}\label{fig:2}
\end{figure}

\begin{figure}
\centering
\includegraphics{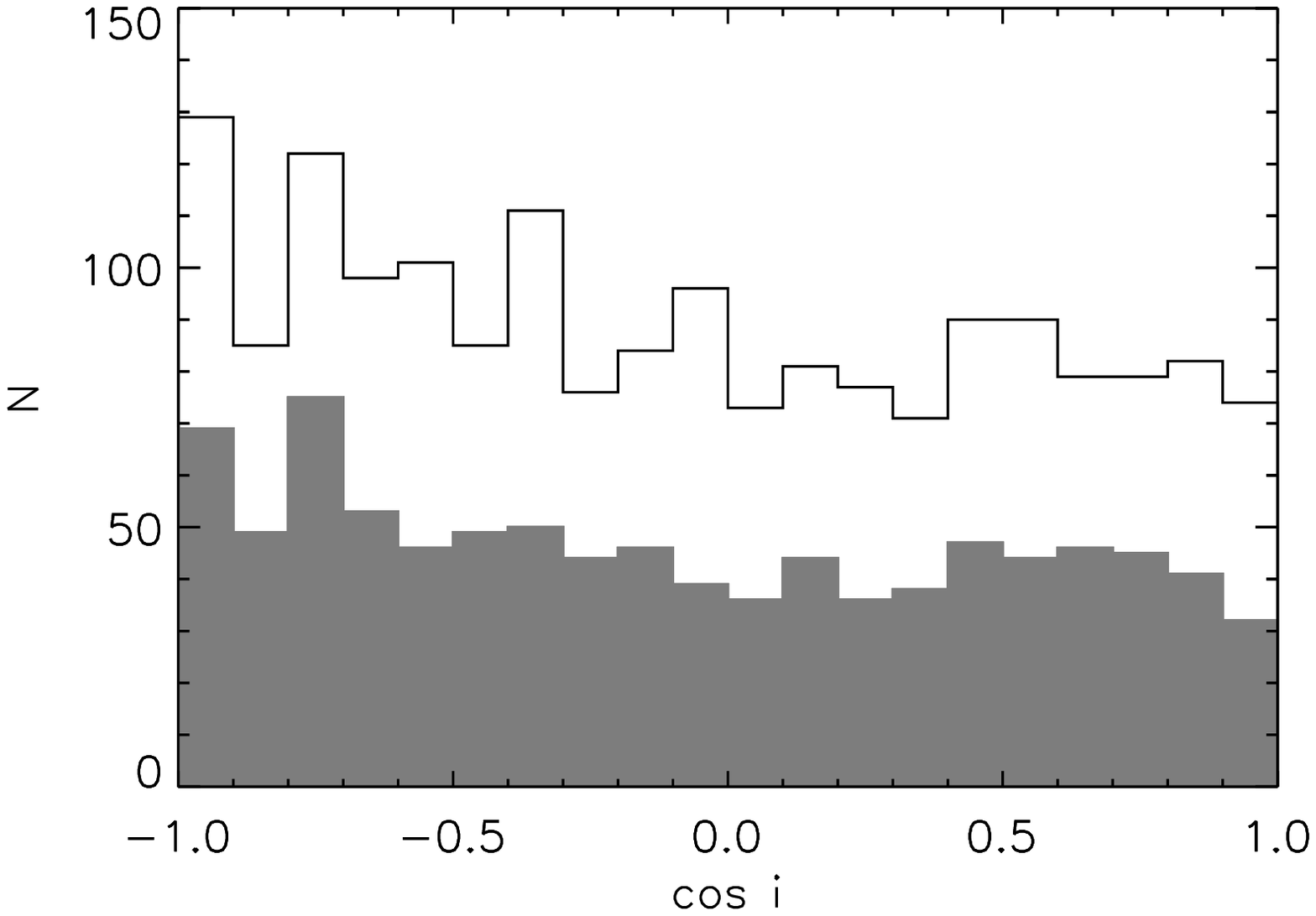}
\caption{Distribution of orbital inclinations with respect to the ecliptic for LPCs as they enter the terrestrial planet region ($q <$ 5 AU). The shaded histogram represents the distribution for only LPCs originating from the inner Oort Cloud, while the unfilled histogram is the distribution for all LPCs.  Inclinations are measured at $r =$ 35 AU immediately after the comet has obtained $q <$ 5 AU.}\label{fig:3}
\end{figure}

\begin{figure}
\centering
\includegraphics{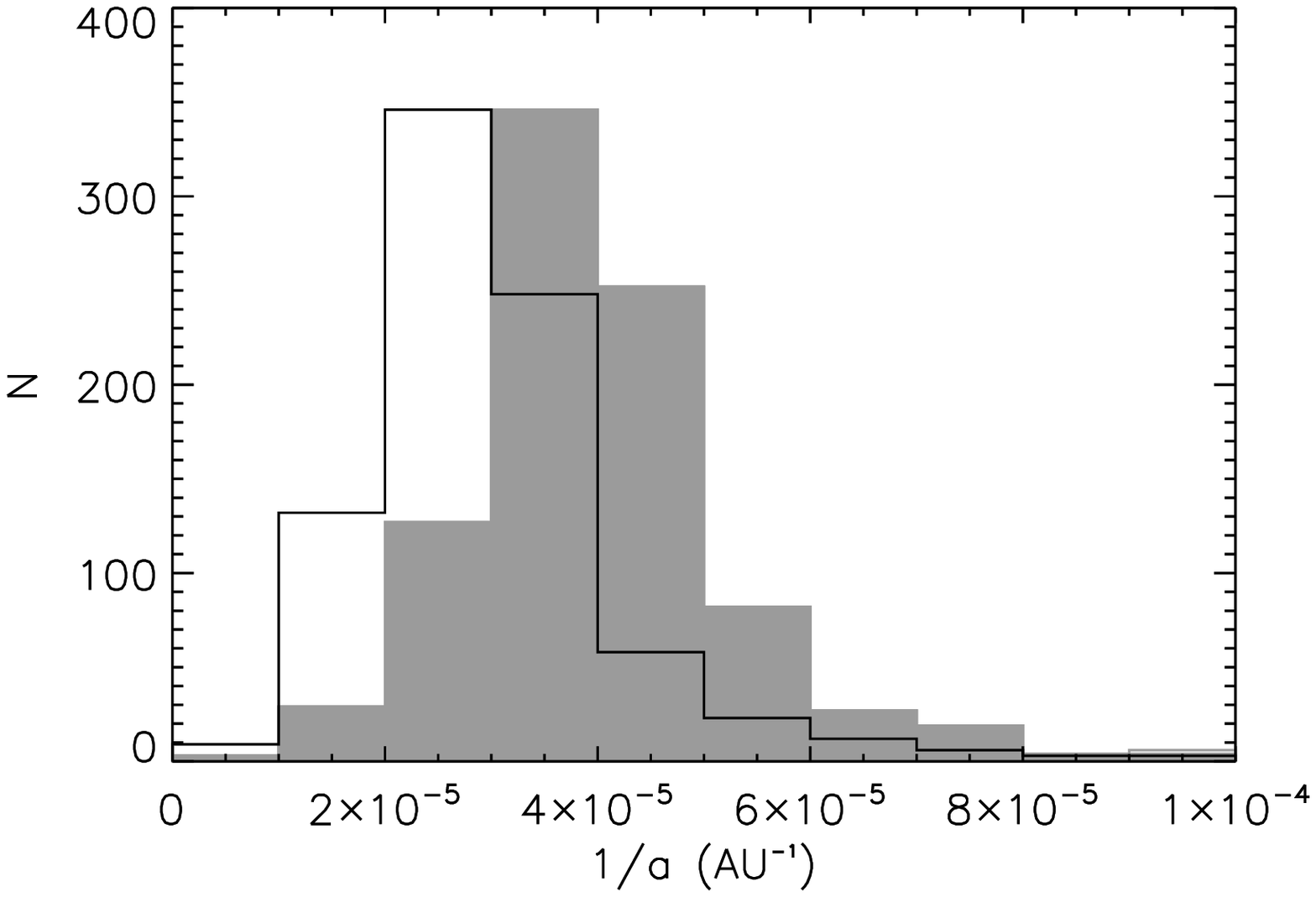}
\caption{Distribution of 1/$a$ for LPCs as they enter the terrestrial planet region ($q <$ 5 AU). The shaded histogram represents the distribution for only LPCs originating from the inner Oort Cloud, while the unfilled histogram is the distribution for LPCs from the outer Oort Cloud.  Semimajor axes are measured at $r =$ 35 AU immediately after the comet has obtained $q <$ 5 AU.}\label{fig:4}
\end{figure}

\begin{figure}
\centering
\includegraphics{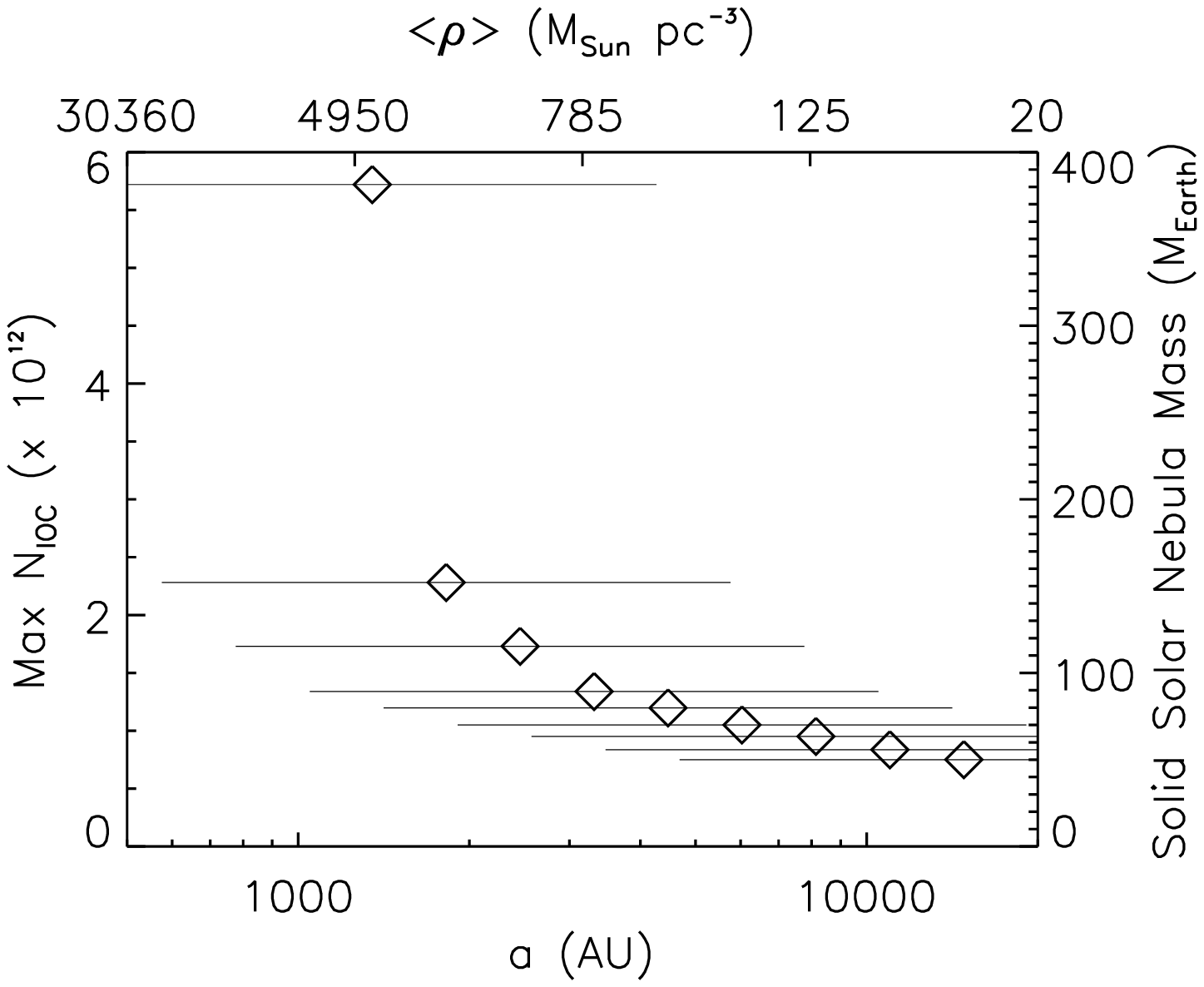}
\caption{A plot of the maximum inner Oort Cloud population (left axis) and the required solar nebula mass in solids (right axis) as function of possible inner Oort Cloud semimajor axis ranges (bottom axis).  The approximate embedded cluster density yielding a given semimajor axis range is shown on the upper axis\cite{bras06}.  The horizontal error bars in this plot mark the entire $a$-range of each hypothetical inner Oort Cloud distribution.  }\label{fig:5}
\end{figure}

\begin{figure}
\centering
\includegraphics{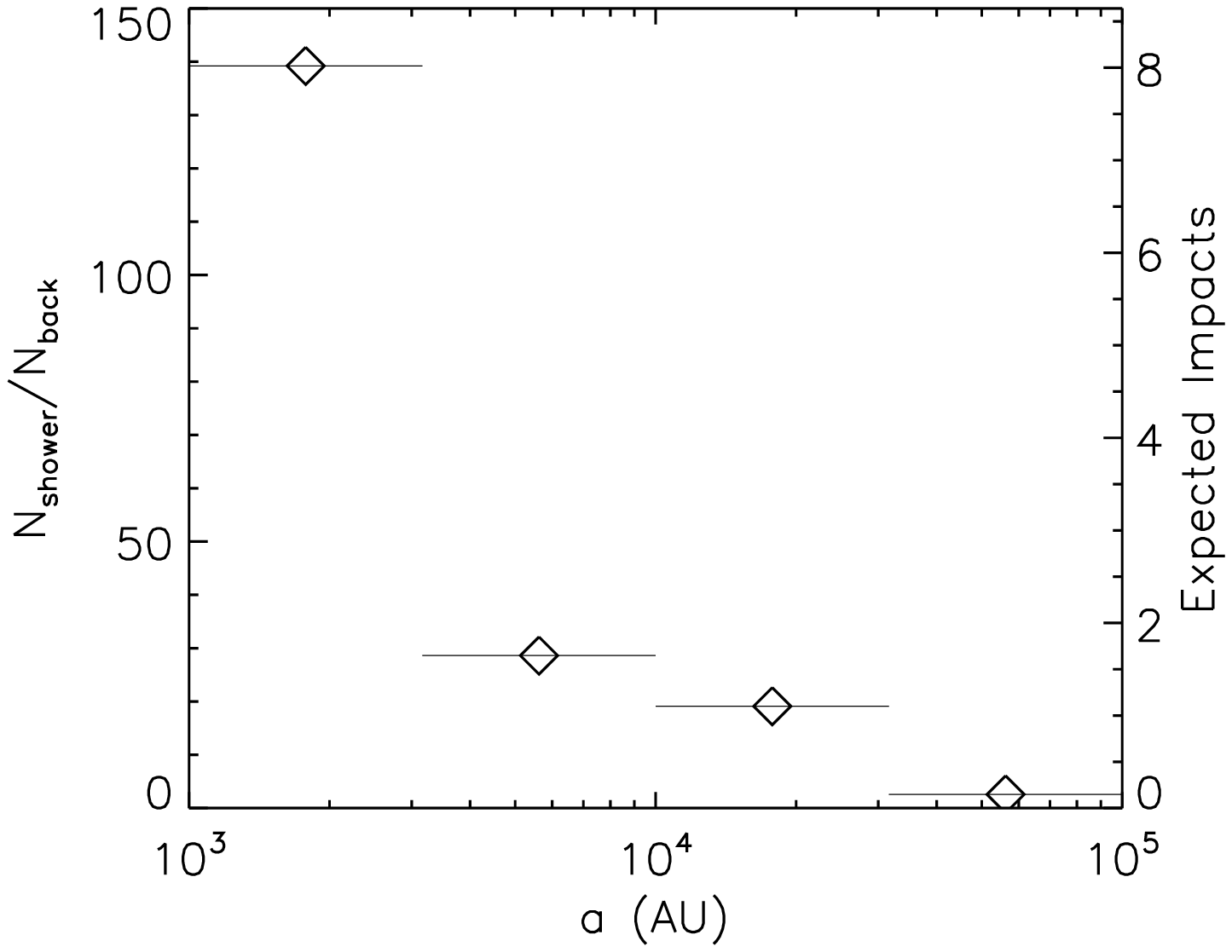}
\caption{A plot of the ratio shower to background LPC production for different ranges of semimajor axis.  On the right axis, the expected impact number is calculated if all Oort Cloud bodies are located in the given semimajor axis range.}\label{fig:6}
\end{figure}

\vspace{3 in}

\bibliographystyle{Science}

\bibliography{maintext}

\begin{scilastnote}
\item This research was funded by a NASA Earth and Space Science Fellowship and an NSF grant (AST-0709191).  Our computing was performed using Purdue Teragrid computing facilities managed with Condor scheduling software (see http://www.cs.wisc.edu/condor).  Allessandro Morbidelli and two anonymous reviewers provided helpful comments that greatly improved our work.
\end{scilastnote}

\end{document}